\documentclass[reprint,
superscriptaddress,
amsmath,amssymb,
aps,
pra,
floatfix,
]{revtex4-2}

\usepackage{graphicx}% Include figure files
\usepackage{dcolumn}% Align table columns on decimal point
\usepackage{bm}% bold math
\usepackage[]{subfigure}
\usepackage{braket}

\begin{document}

\preprint{APS/123-QED}

\title{Dynamical analysis of modal coupling in rare-earth whispering gallery mode micro-lasers}% Force line breaks with \\

\author{Jean-Baptiste Ceppe}
 \altaffiliation[Now at ]{Light-Matter Interaction Unit, OIST Graduate University, Onna, Japan}%Lines break automatically or can be forced with \\
 \email{jean-baptiste.ceppe@oist.jp}
\affiliation{Universit\'{e} de Rennes, CNRS, Institut FOTON - UMR 6082, F-22305 Lannion}%
\author{Patrice F\'{e}ron}
\affiliation{Universit\'{e} de Rennes, CNRS, Institut FOTON - UMR 6082, F-22305 Lannion}%
\author{Michel Mortier}
\affiliation{PSL Research University, Chimie ParisTech - CNRS, Institut de Recherche de Chimie de Paris, Paris 75005, France}%
\author{Yannick Dumeige}%
 \email{yannick.dumeige@univ-rennes1.fr}
\affiliation{Universit\'{e} de Rennes, CNRS, Institut FOTON - UMR 6082, F-22305 Lannion}%

\date{\today}% It is always \today, today,

\begin{abstract}
We report on an experimental study of laser regime in erbium-doped whispering gallery mode (WGM) microspheres under modal-coupling between the co- and counter-propagating modes. The evidence of modal coupling has been observed in the relative intensity noise spectrum of several WGM lasers. Cross-correlation measurements have been carried out in order to determine precisely the emission regimes. It is shown that depending on the material constituting the WGM resonator, frequency locked bidirectional emission or self-modulated regime can be reached. The control of the laser emission regime of WGM micro-lasers is of great importance in the aim of applications in microwave optics or optical sensor miniaturization.
\end{abstract}

\maketitle

%\tableofcontents

\section{\label{Intro}Introduction}

Solid-state whispering gallery mode (WGM) microcavities combine low mode volume and high quality (Q) factors \cite{Vahala03, Chiasera10}. Therefore active WGM microcavities made of materials with optical gain can be used as low threshold lasers \cite{Sandoghdar96}. A lot of WGM lasers \cite{He13} have already been investigated: semiconductor microdisks \cite{McCall92, Zhang15, Elbaz18}, silica microtoroids \cite{Ostby07}, crystalline spheroids \cite{Herr18} or spherical glass microcavities \cite{Lissillour01a} doped with rare-earth ions emitting from visible \cite{Tabataba18} to mid-infrared \cite{Palma16, Palma17, Behzadi17}. These lasers find applications in fundamental physics \cite{Gerard99, Huet2016}, sensing \cite{Ciminelli10, Ward11, Ren17}, optical switching \cite{Liu10} or microwave photonics \cite{Xiao10}. Furthermore, rare-earth doped WGM lasers can reach very narrow linewidth down to 20~kHz in the free running operation \cite{Sandoghdar96, Lissillour01} which gives them many assets for sensing applications \cite{He10} or high-purity microwave generation \cite{Xiao10}. The WGM laser topologies are similar to those of ring cavity lasers, thus, interesting and complex bidirectional dynamical behavior is expected as it is the case for bulk lasers \cite{Schwartz06, Schwartz07, Schwartz08}. Modal coupling mechanisms in solid-state ring laser are well-understood \cite{efanova1968interaction, Khandokhin85}, they have two main contributions: the first is the light reflection or backscattering from optical components whereas the second is due to inhomogeneous gain saturation. The first coupling tends to favor the existence of co- and counter-propagating waves while the other one tends toward unidirectional emission. A competition between theses two opposite effects can lead to a self-modulation regime depending on their relative strength. Furthermore, Rayleigh backscattering in passive high-finesse WGM microcavities is enhanced leading to a strong coupling between co- and counter-propagating modes \cite{Weiss95, Kippenberg2009, Trebaol2010}. Therefore, bidirectional emission may be privileged in solid-state WGM micro-lasers. To date, the dynamics and the coupling of co- and counter-propagating modes have been extensively studied in WGM semiconductor lasers \cite{Sorel03, Giuliani05, Javaloyes09, Liu10} nevertheless this effect has been only rarely addressed in the framework of rare-earth WGM lasers \cite{He10, He10_pra}.
In this paper, we report experimental studies on the dynamical behavior of erbium-doped WGM microsphere lasers in which modal-coupling occurs. In the first part, we briefly remind the different operation regimes of solid-state ring lasers under Rayleigh backscattering. Then, we describe our experimental setup enabling us to measure the relative intensity noise (RIN) and time domain cross-correlations between the co- and counter-propagating WGM laser signals. Finally, we analyze bidirectional laser emissions from WGM micro-lasers made of two different host matrices, namely a phosphate glass (Schott IOG-1 \cite{Ristic16}) and a fluoride glass (ZBLALiP \cite{Rasoloniaina2014}), both doped with erbium ions.

\section{\label{theorie}Laser regimes under Rayleigh backscattering}

\begin{figure}[htbp]
\centering\includegraphics[width=5cm]{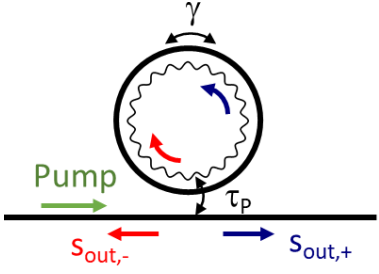}% Here is how to import EPS art
\caption{WGM laser under Rayleigh scattering (coupling rate $\gamma$). The interference pattern of the co- and counter-propagating waves leads to a nonuniform gain saturation which is a nonlinear extra coupling term between the co- and counter-propagating modes. $1/\tau_P$ represents the light-coupling rate from the waveguide to the microcavity.}\label{Fig1}
\end{figure}

Figure \ref{Fig1} shows the sketch of a WGM laser consisting of a microsphere coupled to an access waveguide with a rate $1/\tau_P$. The pump field and the laser emissions $s_{out,+}$ and $s_{out,-}$ are respectively inserted in the cavity and extracted from the cavity thanks to the access line via an evanescent coupling. The Rayleigh backscattering couples the two counter-propagating modes with a rate $\gamma$. Moreover, the interferences between these two modes leads to a periodic gain saturation inducing mode coupling by resonant scattering.
The interaction with the gain medium is well-described using semi-classical approach with Maxwell-Bloch equations considering that the electric field can be written as :
\begin{equation}\label{champE}
E(z,t) = \mathrm{Re} \left[ \tilde{E}_1(t) e^{-j(\omega_0 t - k z)} + \tilde{E}_2 e^{-j(\omega_0 t + k z)}\right]
\end{equation}
where $\omega_0$ is the mean angular frequency of the emitted signal, $z$ is the curvilinear abscissa along the cavity, $k$ is the the wave-vector, $\tilde{E}_{1}$ is the complex amplitude of the counter-clockwise traveling wave ($z>0$) while $\tilde{E}_2$ is the one traveling in the other direction ($z<0$). We note $\tilde{E}_{1,2}=E_{1,2}e^{-j\phi_{1,2}}$ with $E_{1,2}=\left|\tilde{E}_{1,2}\right|$. We assume that the laser is single mode and that $\tilde{E}_1$ and $\tilde{E}_2$ have the same polarization. The erbium-doped medium is represented by a three-level system with an excited lifetime $T_1 = 1/A$ (where $A$ is the spontaneous emission rate) and an emission cross section $\sigma$. In the case of co- and counter-propagating waves, the two considered fields have the same mode volume and photon lifetime $\tau_P$ \cite{Kippenberg2009}. $N$ is the refractive index of the cavity and $L = 2\pi a$ is the length of the cavity (which represents also the length of the active medium) where $a$ is the radius of the resonator. As we deal with erbium-doped medium, the polarization can be adiabatically suppressed and the whole system is described only by the evolution of the electric fields amplitudes ${E}_{1,2}$ and the inverted ion density $\mathcal{N}$. The existence of two lasing modes leads to an interference pattern, modulating the gain medium, as shown in Fig.~\ref{Fig1}. This diffraction grating induces a supplementary nonlinear coupling. Under the hypothesis of a slightly above threshold pumping rate, the whole system is described by: 
\begin{eqnarray}\label{EBO}
\frac{\mathrm{d} {E}_{1,2}}{\mathrm{d}t} &=&  -\frac{1}{2\tau_{P}}{E}_{1,2} \mp \frac{\gamma}{2}{E}_{2,1}\sin\left(\Phi + \theta_{1,2}\right) \\
& & + \frac{c\sigma}{2NL}\oint\mathcal{N}\left[{E}_{1,2} + {E}_{2,1} \cos\left(2kz+\Phi\right)\right]\mathrm{d}z\nonumber   \\
\frac{\mathrm{d}\Phi}{\mathrm{d}t} &=& -\frac{\gamma}{2}\left[ \frac{E_2}{E_1}\cos\left(\Phi + \theta_1\right) - \frac{E_1}{E_2}\cos\left(\Phi + \theta_2\right) \right] \\
& & - \frac{c\sigma}{2NL}\left( \frac{E_1^2 + E_2^2}{E_1E_2}\right) \oint \mathcal{N}\sin\left( 2kz + \Phi\right) \mathrm{d}z\nonumber\\
\frac{\partial \mathcal{N}}{\partial t} &=&  \frac{\eta\mathcal{N}_{th}-\mathcal{N}}{\tau}-\frac{\mathcal{N}}{\tau} \frac{\left| \tilde{E}_1 + \tilde{E}_2 \right|^2 }{E_\text{sat}^2 }
%\frac{\mathrm{d} \mathcal{N}_{1}}{\mathrm{d} t} &=&- \frac{\mathcal{N}_1 }{\tau} -\frac{\mathcal{N}_{th}}{\tau} \left(\frac{\tilde{E}_1\tilde{E}_2^{*}}{E_\text{sat}^2}\right)
\end{eqnarray}
where $\Phi=\phi_1-\phi_2$, $\mathcal{N}_{th}$ is the inverted ion density at threshold, $\eta$ is the excitation rate, $\tau$ is the recovery time ($\tau^{-1} = W_P + A$, $W_P$ being the pumping rate) \cite{siegman1986lasers} and $\theta_{1,2}$ is the acquired backscattering phase (we assume that $\theta_1$ and $\theta_2$ are close to each other and define $2\delta_\theta = \theta_1 - \theta_2 \ll 1$). One can see in Eq. (\ref{EBO}) that $\gamma$ is a linear coupling term while $\mathcal{N}$ induces both gain and nonlinear coupling between $E_1$ and $E_2$. Theses two coupling terms, and their relative strength, drive the laser to different regimes. From a stability analysis, it is possible to derive  that the laser operates at three different regimes defined by the value of $\gamma$ \cite{schwartz2006gyrolaser}. Under our hypothesis, the laser exhibits two steady-states (unidirectional and bidirectional regimes) and a permanent state (self-modulation). \textit{i) The unidirectional regime} is obtained for low values of $\gamma$ such as $\gamma < \frac{\omega_{r}}{2}$, where $\omega_{r}$ is the relaxation oscillation angular frequency \cite{siegman1986lasers}. In this regime, the backscattering is not high enough to authorize the existence of two traveling modes and the mode with higher amplitude gets the higher gain \cite{perevedentseva1980theory}. \textit{ii) The self-modulated regime} is obtained for intermediate values of $\gamma$. This is a periodic permanent regime in which the two lasing modes are anti-phase oscillating at the angular frequency $\gamma$ \cite{Schwartz07}. The modulation contrast $\mathcal{C}$ decreases with $\gamma$
\begin{equation}
\mathcal{C} = \sqrt{1 - \left( \frac{2\gamma\left| \delta_\theta\right|\tau_P'}{\eta - 1 - \gamma\left| \delta_\theta\right|\tau_P'}\right)^2}
\end{equation}
with $\tau_P'^{-1} = \tau_P^{-1} - \gamma\left| \delta_\theta\right|$. When $\mathcal{C}$ reaches zero, the laser shows a bidirectional behavior. \textit{iii) This bidirectional regime} is obtained for high values of $\gamma$ \cite{Ceppe18}
\begin{equation}\label{transition}
\gamma > \frac{1}{\tau_P \left| \delta_\theta\right|}\left(\frac{\eta-1}{\eta+2}\right),
\end{equation}
in this case, the two traveling waves are automatically frequency-locked together \cite{siegman1986lasers}, which is the main problem for solid-state ring laser gyroscopes \cite{Schwartz06}.

\section{\label{experience}Experiments}

\subsection{Setup description}\label{setup}

We have two glasses with different host matrix at our disposal: IOG-1 an industrial phosphate glass from Schott and ZBLALiP a fluoride glass. Both glasses are doped with erbium ions ($3.25\times10^{20}$~ions/cm$^3$ for IOG-1 and $0.2\times10^{20}$~ions/cm$^3$ for ZBLALiP).
\begin{figure*}[htbp]
\includegraphics[width=17cm]{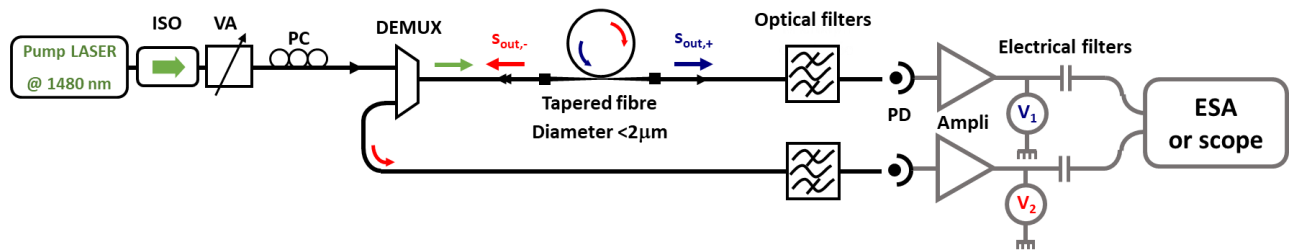}
\caption{Experimental setup. ISO: optical isolator, VA: variable attenuator, PC: polarization controller, DEMUX: wavelength demultiplexer, PD: photodiodes, Ampli: transimpedance amplifiers, Optical filter: narrow bandpass optical filters, Electrical filters: DC block filters or high-pass filters, ESA: electrical spectrum analyzer. The ESA is used to measure the RIN in the frequency domain whereas the oscilloscope is used for cross-correlation measurements in the time domain.}\label{Fig2}
\end{figure*}
Microspheres are manufactured by a melting technique from a glass powder using a plasma torch \cite{Rasoloniaina2014}. This process enables us to make microspheres with diameters $2a$ between $60$ and $150$ $\mu$m. Figure \ref{Fig2} shows the experimental setup used to analyze the laser emission of our microcavities. The erbium transition $^{4}I_{13/2}\rightarrow{^{4}I_{15/2}}$ is excited using a $1480~\mathrm{nm}$ laser diode (pump laser) whose maximal power is 100~mW. We use a tapered fiber, whose diameter is reduced to about 1 $\mu$m, to couple the pump into the microcavity.
\begin{figure*}[htbp]
\includegraphics[width=17cm]{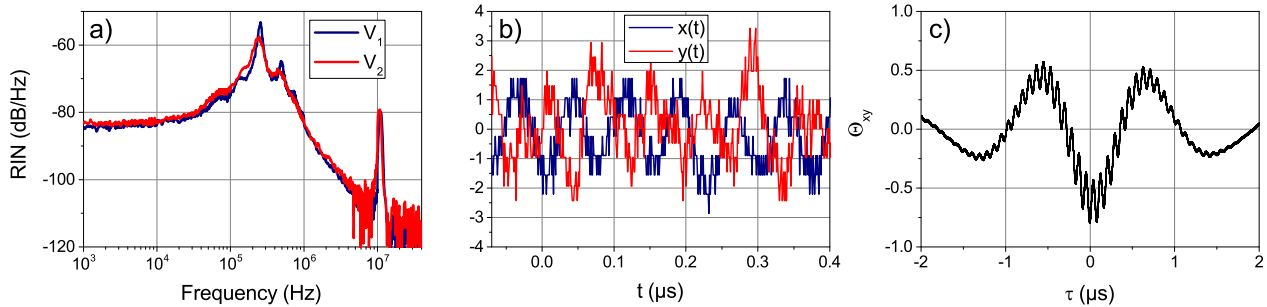}
\caption{Experimental results obtained in a 105~$\mu$m diameter IOG-1 microsphere. a) RIN spectrum for the co- and counter-propagating emitted modes. b) Time domain series $x(t)$ and $y(t)$ obtained from centering and normalization of $V_1(t)$ and $V_2(t)$. c) Normalized cross-correlation function.}\label{Fig3}
\end{figure*}
The co- and counter-propagating signals are collected using the same tapered fiber \cite{Lissillour01a, Ceppe17}. $s_{out,-}$ is separated from the initial pump with a wavelength demultiplexer. The laser emissions of interest are then optically filtered using manually tunable optical filters (Yenista XTM-50, bandwidth 4~GHz). RIN measurements are performed on both emission directions using a photodetection system already reported in previous work \cite{Ceppe17} and depicted in Fig.~\ref{Fig2}. The low-noise transimpedance amplifier (Femto DHPCA-100) converts the photocurrents into voltages $V_1$ and $V_2$. Electrical filters (Thorlabs EF500, cut-off frequency 1 Hz) are used to suppress the DC component of the two voltages. Power spectrum density of the noise is measured using an electrical spectrum analyzer (ESA) (Agilent, bandwidth : 40~MHz). The whole photo-detecting chain is well suited to RIN measuring up to 10-20~MHz. By replacing the ESA by an oscilloscope (Wavemaster LeCroy) we can also perform the photodetection in the time domain and thus infer cross-correlations between the counter-propagating laser emissions using the following procedure. First, voltages $V_1$ and $V_2$ are normalized as follows
\begin{eqnarray}
x(t) &=& \frac{V_1(t) - \braket{V_1}}{\sigma_{V_1}}\\
y(t) &=& \frac{V_2(t)- \braket{V_2}}{\sigma_{V_2}},
\end{eqnarray}
where $\braket{\cdot}$ is the temporal average and $\sigma_{V_{1,2}}$ the standard deviation of $V_{1,2}$. We then calculate the cross-correlation function $\Gamma_{xy}$ and the auto-correlation functions $\Gamma_{xx}$ and $\Gamma_{yy}$ defined by
\begin{eqnarray}
\Gamma_{xy}(\tau) &=& \braket{x(t)y(t-\tau)}\\
\Gamma_{xx}(\tau) &=& \braket{x(t)x(t-\tau)}\\
\Gamma_{yy}(\tau) &=& \braket{y(t)y(t-\tau)}.
\end{eqnarray}
Finally, we obtain the normalized cross-correlation function
\begin{equation}
\Theta_{xy}(\tau) = \frac{\Gamma_{xy}(\tau)}{\sqrt{\Gamma_{xx}(0)\Gamma_{yy}(0)}}.
\end{equation}

\subsection{Experimental results in IOG-1 glass}\label{results1}

We first used a IOG-1 microsphere with a diameter of 105~$\mu$m. The optical filters are both centered on the laser emission wavelength which is the same for both propagation directions (1564 nm). After filtering, the photo-detected powers are 150 nW for $s_{out,+}$ and 86 nW for $s_{out,-}$. RIN spectra for co- and counter-propagating modes are presented in Fig. \ref{Fig3}.a).
\begin{figure*}[htbp]
\includegraphics[width=17cm]{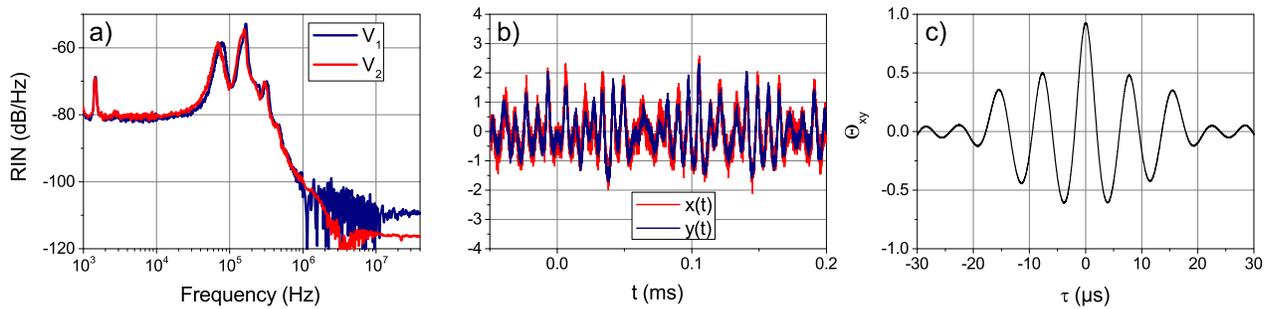}% Here is how to import EPS art
\caption{Experimental results obtained in a 100~$\mu$m diameter ZBLALiP microsphere. a) RIN spectrum for the co- and counter-propagating emitted modes. b) Time domain series $x(t)$ and $y(t)$ obtained from centering and normalization of $V_1(t)$ and $V_2(t)$. c) Normalized cross-correlation function.}\label{Fig4}
\end{figure*}
These spectra show one peak at the relaxation oscillation frequency which is the signature of a class-B laser operation and also some associated harmonic frequencies due to the low mode volume of the cavity \cite{Ceppe17}. The RIN spectra of the two counter-propagating modes are similar, in particular they have the same relaxation oscillation frequencies ($\frac{\omega_r}{2\pi}=257~\mathrm{kHz}$) which shows that the two modes have the same photon lifetimes and pumping rates. Furthermore since the relative intensities of the harmonic and fundamental peaks are the same, the two counter-propagating modes have the same mode volume \cite{Ceppe17}. Therefore we can conclude that the WGM $s_{out,+}$ and $s_{out,-}$ have the same radial ($n$) and orbital ($\ell$) orders and opposite azimutal ($m$) order since they propagate in opposite directions. There is also an additional RIN peak at 12~MHz. This RIN peak can result from the self-modulation behavior of the WGM laser submitted to modal-coupling owing to Rayleigh backscattering as recalled in section \ref{theorie}. To confirm this hypothesis we also performed cross-correlation measurements following the procedure presented in \ref{setup}. As we are only interested into high frequencies (higher than the relaxation oscillation frequency), and because of the high intrinsic noise level of the laser itself, we use an electrical high-pass filter with a cut-off frequency of 600~kHz (Thorlabs EF517). Normalized voltages $x(t)$ and $y(t)$ and the cross-correlation function $\Theta_{xy}(\tau)$ are presented in Fig.~\ref{Fig3}.b) and \ref{Fig3}.c).
\begin{figure}
\includegraphics[width=7cm]{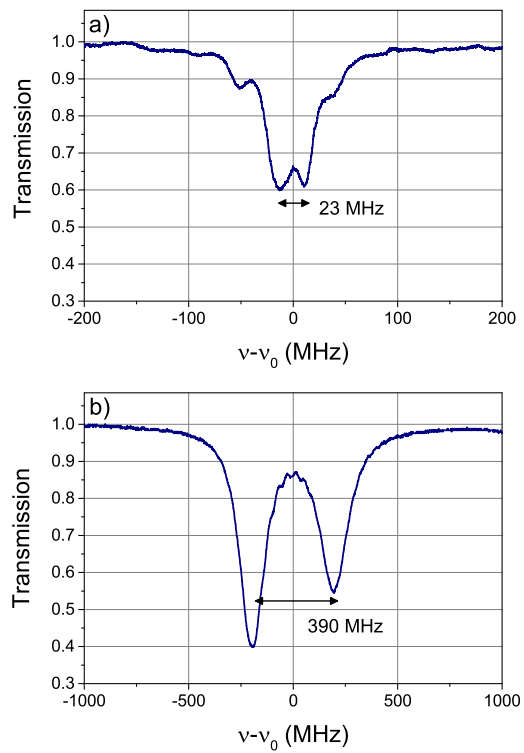}% Here is how to import EPS art
\caption{Typical linear transmission spectrum measured for a) a IOG-1 microsphere under the laser threshold, b) a ZBLALiP passive (without erbium doping) microsphere. $\nu$ is the frequency of the probe laser and $\nu_0$ is the resonance frequency of the WGM microsphere.}\label{Fig5}
\end{figure}
The phase opposition between $x(t)$ and $y(t)$, clearly seen in Fig. \ref{Fig3}.b), leads to a strong anti-correlation, highlighted by the value of the cross-correlation for short times $\Theta_{xy}(\tau=0)=-0.8$. This confirms the hypothesis of self-modulation operation of our IOG-1 WGM laser due to the competition between Rayleigh backscattering and gain saturation (see section \ref{theorie}). The over-modulation with a long period (corresponding to the third harmonic of the relaxation oscillation frequency) appearing in $\Theta_{xy}(\tau)$ comes from an excess of intrinsic noise that has not been suppressed by the electrical filtering. Note that the cross-correlation function associated to this modulation, given by the envelope of the curve shown in Fig. \ref{Fig3}.c), has a strong negative value for short delays as well.

\subsection{Experimental results in ZBLALiP glass}
We also tested a ZBLALiP microsphere with a diameter of 100 $\mu$m. After optical filtering at 1558~nm, the photo-detected powers are 157~nW for $s_{out,+}$ and 27~nW for $s_{out,-}$. We observe two relaxation oscillation frequencies in the RIN spectra due to the fact that for a given emission direction the WGM has now two modes. Nevertheless, the two RIN spectra for the two opposite propagation directions perfectly match and thus the signals emitted in both direction correspond to the same WGM. As our RIN measurement setup is limited to frequency less than 40~MHz we cannot conclude on the laser operation regime since the beating frequency could appear at higher frequencies. Cross-correlation measurements are performed using a DC block electrical filter. Normalized voltages $x(t)$ and $y(t)$ and normalized cross-correlation function $\Theta_{xy}(\tau)$ are presented in Fig.~\ref{Fig4}.b) and \ref{Fig4}.c). Now we observe a perfect correlation between the co- and counter-propagating signals. Under the assumption that the correlations between the two counter-propagating signals persist at the relaxation oscillation frequency and its harmonics (see section \ref{results1}) we can conclude that the laser operates in the bidirectional regime. 

\subsection{Discussion}

The two studied WGM lasers emit light in both direction but operate in two different regimes. The IOG-1 laser for which the co- and counter-propagating modes are emitted in anti-phase operates in a self-modulation regime. In contrast, the fluoride glass WGM laser shows two emitted signals in phase, this can be attributed to a frequency-locked bidirectional emission. Inspection of Eq. (\ref{transition}) reveals that one of the key parameters determining the laser operation is the linear coupling rate between the two counter-propagating modes. Assuming that photon lifetimes and pumping rates are comparable, we can deduce from the previous experiments that the linear coupling strength is much stronger for ZBLALiP than for IOG-1 glass. This can be confirmed by linear optical transmission measurements. This consists in using a tunable narrow-band laser ($\le 150~\mathrm{kHz}$) frequency-swept across the WGM resonance and simultaneously recording the transmission \cite{Dumeige08, Rasoloniaina2014}. A thorough frequency calibration allows to obtain the linear transmission spectrum. We give such measurements done for our two glasses in Fig. \ref{Fig5}. Note that these spectra have been obtained with other microspheres made of the same materials as those used in laser experiments. Nevertheless it informs us on the typical frequency splitting $2\delta_C\approx\frac{\gamma}{2\pi}$ \cite{Trebaol2010} which can be reached both for IOG-1 and ZBLALiP WGM microspheres. For IOG-1 the frequency splitting is such that $2\delta_C=$23~MHz whereas for ZBLALiP it is around 390~MHz. This confirms that the Rayleigh backscattering is much stronger in the fluoride glass than in the phosphate glass. Moreover the frequency splitting obtained for IOG-1 WGM resonators is comparable to the beating frequency (12~MHz) measured in the RIN when the self-modulation operation is reached. Finally we can note that in both cases the coupling rate is much larger than the relaxation oscillation frequency which supports the fact that these WGM lasers do not operate in the unidirectional regime.

\section{Conclusion}

We have simultaneously measured the RIN spectra of the two counter-propagating signals emitted by erbium doped WGM class-B micro-lasers. In addition to noise peaks at the relaxation oscillation frequency and its harmonics, the analysis of the RIN spectrum of phosphate glass microspherical laser reveals an extra peak due to the self-modulation operation induced by a Rayleigh backscattering induced modal coupling. Cross-correlation measurements between the co- and counter-propagating laser signals show a neat anti-phase oscillation confirming the self-modulation regime \cite{Ceppe17}. We have repeated this measurement using a fluoride glass WGM laser and in this case a bidirectional operation, confirmed by a cross-correlation measurement, has been observed. This behavior difference is well supported by the fact that Rayleigh backscattering is much stronger in our fluoride glass microspheres than in phosphate glass microcavities. Controlling the bidirectional feature of miniaturized WGM lasers is crucial for applications. For instance, integrated laser gyroscopes must operate in the self-modulation regime \cite{Schwartz07}. Furthermore, a good knowledge of the physical and optical properties of materials would be of importance to properly design the micro-laser gyroscope. When the modal coupling rate has been experimentally determined, the photon lifetime, which can be controlled via the evanescent coupling with the access line, can be adapted to reach the self-modulation regime using Eq. (\ref{transition}). Conversely, for all-optical compact microwave-sources, a bidirectional behavior is preferred since it would avoid the apparition of spurious spikes in the RF spectrum.

\begin{acknowledgments}
J.-B. Ceppe thanks  the Centre National d'\'{E}tudes Spatiales (CNES) and R\'{e}gion Bretagne (ARED) for financial support. Y. Dumeige is member of the Institut Universitaire de France. The authors acknowledge fruitful discussions with S. Trebaol.
\end{acknowledgments}


\begin{thebibliography}{44}%
\makeatletter
\providecommand \@ifxundefined [1]{%
 \@ifx{#1\undefined}
}%
\providecommand \@ifnum [1]{%
 \ifnum #1\expandafter \@firstoftwo
 \else \expandafter \@secondoftwo
 \fi
}%
\providecommand \@ifx [1]{%
 \ifx #1\expandafter \@firstoftwo
 \else \expandafter \@secondoftwo
 \fi
}%
\providecommand \natexlab [1]{#1}%
\providecommand \enquote  [1]{``#1''}%
\providecommand \bibnamefont  [1]{#1}%
\providecommand \bibfnamefont [1]{#1}%
\providecommand \citenamefont [1]{#1}%
\providecommand \href@noop [0]{\@secondoftwo}%
\providecommand \href [0]{\begingroup \@sanitize@url \@href}%
\providecommand \@href[1]{\@@startlink{#1}\@@href}%
\providecommand \@@href[1]{\endgroup#1\@@endlink}%
\providecommand \@sanitize@url [0]{\catcode `\\12\catcode `\$12\catcode
  `\&12\catcode `\#12\catcode `\^12\catcode `\_12\catcode `\%12\relax}%
\providecommand \@@startlink[1]{}%
\providecommand \@@endlink[0]{}%
\providecommand \url  [0]{\begingroup\@sanitize@url \@url }%
\providecommand \@url [1]{\endgroup\@href {#1}{\urlprefix }}%
\providecommand \urlprefix  [0]{URL }%
\providecommand \Eprint [0]{\href }%
\providecommand \doibase [0]{https://doi.org/}%
\providecommand \selectlanguage [0]{\@gobble}%
\providecommand \bibinfo  [0]{\@secondoftwo}%
\providecommand \bibfield  [0]{\@secondoftwo}%
\providecommand \translation [1]{[#1]}%
\providecommand \BibitemOpen [0]{}%
\providecommand \bibitemStop [0]{}%
\providecommand \bibitemNoStop [0]{.\EOS\space}%
\providecommand \EOS [0]{\spacefactor3000\relax}%
\providecommand \BibitemShut  [1]{\csname bibitem#1\endcsname}%
\let\auto@bib@innerbib\@empty
%</preamble>
\bibitem [{\citenamefont {Vahala}(2003)}]{Vahala03}%
  \BibitemOpen
  \bibfield  {author} {\bibinfo {author} {\bibfnamefont {K.~J.}\ \bibnamefont
  {Vahala}},\ }\bibfield  {title} {\bibinfo {title} {Optical microcavities},\
  }\href@noop {} {\bibfield  {journal} {\bibinfo  {journal} {Nature}\ }\textbf
  {\bibinfo {volume} {424}},\ \bibinfo {pages} {839} (\bibinfo {year}
  {2003})}\BibitemShut {NoStop}%
\bibitem [{\citenamefont {{Ilchenko}}\ and\ \citenamefont
  {{Matsko}}(2006)}]{Ilchenko06}%
  \BibitemOpen
  \bibfield  {author} {\bibinfo {author} {\bibfnamefont {V.~S.}\ \bibnamefont
  {{Ilchenko}}}\ and\ \bibinfo {author} {\bibfnamefont {A.~B.}\ \bibnamefont
  {{Matsko}}},\ }\bibfield  {title} {\bibinfo {title} {Optical resonators with
  whispering-gallery modes-part {II}: applications},\ }\href
  {https://doi.org/10.1109/JSTQE.2005.862943} {\bibfield  {journal} {\bibinfo
  {journal} {IEEE J. Sel. Topics Quantum Electron.}\ }\textbf {\bibinfo
  {volume} {12}},\ \bibinfo {pages} {15} (\bibinfo {year} {2006})}\BibitemShut
  {NoStop}%
\bibitem [{\citenamefont {Chiasera}\ \emph {et~al.}(2010)\citenamefont
  {Chiasera}, \citenamefont {Dumeige}, \citenamefont {F\'{e}ron}, \citenamefont
  {Ferrari}, \citenamefont {Jestin}, \citenamefont {Nunzi~Conti}, \citenamefont
  {Pelli}, \citenamefont {Soria},\ and\ \citenamefont {Righini}}]{Chiasera10}%
  \BibitemOpen
  \bibfield  {author} {\bibinfo {author} {\bibfnamefont {A.}~\bibnamefont
  {Chiasera}}, \bibinfo {author} {\bibfnamefont {Y.}~\bibnamefont {Dumeige}},
  \bibinfo {author} {\bibfnamefont {P.}~\bibnamefont {F\'{e}ron}}, \bibinfo
  {author} {\bibfnamefont {M.}~\bibnamefont {Ferrari}}, \bibinfo {author}
  {\bibfnamefont {Y.}~\bibnamefont {Jestin}}, \bibinfo {author} {\bibfnamefont
  {G.}~\bibnamefont {Nunzi~Conti}}, \bibinfo {author} {\bibfnamefont
  {S.}~\bibnamefont {Pelli}}, \bibinfo {author} {\bibfnamefont
  {S.}~\bibnamefont {Soria}},\ and\ \bibinfo {author} {\bibfnamefont
  {G.}~\bibnamefont {Righini}},\ }\bibfield  {title} {\bibinfo {title}
  {Spherical whispering-gallery-mode microresonators},\ }\href
  {https://doi.org/10.1002/lpor.200910016} {\bibfield  {journal} {\bibinfo
  {journal} {Laser Photonics Rev.}\ }\textbf {\bibinfo {volume} {4}},\ \bibinfo
  {pages} {457} (\bibinfo {year} {2010})}\BibitemShut {NoStop}%
\bibitem [{\citenamefont {Sandoghdar}\ \emph {et~al.}(1996)\citenamefont
  {Sandoghdar}, \citenamefont {Treussart}, \citenamefont {Hare}, \citenamefont
  {Lef\`evre-Seguin}, \citenamefont {Raimond},\ and\ \citenamefont
  {Haroche}}]{Sandoghdar96}%
  \BibitemOpen
  \bibfield  {author} {\bibinfo {author} {\bibfnamefont {V.}~\bibnamefont
  {Sandoghdar}}, \bibinfo {author} {\bibfnamefont {F.}~\bibnamefont
  {Treussart}}, \bibinfo {author} {\bibfnamefont {J.}~\bibnamefont {Hare}},
  \bibinfo {author} {\bibfnamefont {V.}~\bibnamefont {Lef\`evre-Seguin}},
  \bibinfo {author} {\bibfnamefont {J.~M.}\ \bibnamefont {Raimond}},\ and\
  \bibinfo {author} {\bibfnamefont {S.}~\bibnamefont {Haroche}},\ }\bibfield
  {title} {\bibinfo {title} {Very low threshold whispering-gallery-mode
  microsphere laser},\ }\href {https://doi.org/10.1103/PhysRevA.54.R1777}
  {\bibfield  {journal} {\bibinfo  {journal} {Phys. Rev. A}\ }\textbf {\bibinfo
  {volume} {54}},\ \bibinfo {pages} {R1777} (\bibinfo {year}
  {1996})}\BibitemShut {NoStop}%
\bibitem [{\citenamefont {He}\ \emph {et~al.}(2013)\citenamefont {He},
  \citenamefont {\"{O}zdemir},\ and\ \citenamefont {Yang}}]{He13}%
  \BibitemOpen
  \bibfield  {author} {\bibinfo {author} {\bibfnamefont {L.}~\bibnamefont
  {He}}, \bibinfo {author} {\bibfnamefont {a.~K.}\ \bibnamefont
  {\"{O}zdemir}},\ and\ \bibinfo {author} {\bibfnamefont {L.}~\bibnamefont
  {Yang}},\ }\bibfield  {title} {\bibinfo {title} {Whispering gallery
  microcavity lasers},\ }\href {https://doi.org/10.1002/lpor.201100032}
  {\bibfield  {journal} {\bibinfo  {journal} {Laser Photonics Rev.}\ }\textbf
  {\bibinfo {volume} {7}},\ \bibinfo {pages} {60} (\bibinfo {year}
  {2013})}\BibitemShut {NoStop}%
\bibitem [{\citenamefont {McCall}\ \emph {et~al.}(1992)\citenamefont {McCall},
  \citenamefont {Levi}, \citenamefont {Slusher}, \citenamefont {Pearton},\ and\
  \citenamefont {Logan}}]{McCall92}%
  \BibitemOpen
  \bibfield  {author} {\bibinfo {author} {\bibfnamefont {S.~L.}\ \bibnamefont
  {McCall}}, \bibinfo {author} {\bibfnamefont {A.~F.~J.}\ \bibnamefont {Levi}},
  \bibinfo {author} {\bibfnamefont {R.~E.}\ \bibnamefont {Slusher}}, \bibinfo
  {author} {\bibfnamefont {S.~J.}\ \bibnamefont {Pearton}},\ and\ \bibinfo
  {author} {\bibfnamefont {R.~A.}\ \bibnamefont {Logan}},\ }\bibfield  {title}
  {\bibinfo {title} {Whispering--gallery mode microdisk lasers},\ }\href
  {https://doi.org/10.1063/1.106688} {\bibfield  {journal} {\bibinfo  {journal}
  {Appl. Phys. Lett}\ }\textbf {\bibinfo {volume} {60}},\ \bibinfo {pages}
  {289} (\bibinfo {year} {1992})}\BibitemShut {NoStop}%
\bibitem [{\citenamefont {Zhang}\ \emph {et~al.}(2015)\citenamefont {Zhang},
  \citenamefont {Zhang}, \citenamefont {Li}, \citenamefont {Cheung},
  \citenamefont {Feng},\ and\ \citenamefont {Choi}}]{Zhang15}%
  \BibitemOpen
  \bibfield  {author} {\bibinfo {author} {\bibfnamefont {Y.}~\bibnamefont
  {Zhang}}, \bibinfo {author} {\bibfnamefont {X.}~\bibnamefont {Zhang}},
  \bibinfo {author} {\bibfnamefont {K.~H.}\ \bibnamefont {Li}}, \bibinfo
  {author} {\bibfnamefont {Y.~F.}\ \bibnamefont {Cheung}}, \bibinfo {author}
  {\bibfnamefont {C.}~\bibnamefont {Feng}},\ and\ \bibinfo {author}
  {\bibfnamefont {H.~W.}\ \bibnamefont {Choi}},\ }\bibfield  {title} {\bibinfo
  {title} {Advances in {III}-nitride semiconductor microdisk lasers},\ }\href
  {https://doi.org/10.1002/pssa.201431745} {\bibfield  {journal} {\bibinfo
  {journal} {Phys. Status Solidi A}\ }\textbf {\bibinfo {volume} {212}},\
  \bibinfo {pages} {960} (\bibinfo {year} {2015})}\BibitemShut {NoStop}%
\bibitem [{\citenamefont {Elbaz}\ \emph {et~al.}(2018)\citenamefont {Elbaz},
  \citenamefont {El~Kurdi}, \citenamefont {Aassime}, \citenamefont {Sauvage},
  \citenamefont {Checoury}, \citenamefont {Sagnes}, \citenamefont {Baudot},
  \citenamefont {Boeuf},\ and\ \citenamefont {Boucaud}}]{Elbaz18}%
  \BibitemOpen
  \bibfield  {author} {\bibinfo {author} {\bibfnamefont {A.}~\bibnamefont
  {Elbaz}}, \bibinfo {author} {\bibfnamefont {M.}~\bibnamefont {El~Kurdi}},
  \bibinfo {author} {\bibfnamefont {A.}~\bibnamefont {Aassime}}, \bibinfo
  {author} {\bibfnamefont {S.}~\bibnamefont {Sauvage}}, \bibinfo {author}
  {\bibfnamefont {X.}~\bibnamefont {Checoury}}, \bibinfo {author}
  {\bibfnamefont {I.}~\bibnamefont {Sagnes}}, \bibinfo {author} {\bibfnamefont
  {C.}~\bibnamefont {Baudot}}, \bibinfo {author} {\bibfnamefont
  {F.}~\bibnamefont {Boeuf}},\ and\ \bibinfo {author} {\bibfnamefont
  {P.}~\bibnamefont {Boucaud}},\ }\bibfield  {title} {\bibinfo {title}
  {Germanium microlasers on metallic pedestals},\ }\href
  {https://doi.org/10.1063/1.5025705} {\bibfield  {journal} {\bibinfo
  {journal} {APL Photonics}\ }\textbf {\bibinfo {volume} {3}},\ \bibinfo
  {pages} {106102} (\bibinfo {year} {2018})}\BibitemShut {NoStop}%
\bibitem [{\citenamefont {Ostby}\ \emph {et~al.}(2007)\citenamefont {Ostby},
  \citenamefont {Yang},\ and\ \citenamefont {Vahala}}]{Ostby07}%
  \BibitemOpen
  \bibfield  {author} {\bibinfo {author} {\bibfnamefont {E.~P.}\ \bibnamefont
  {Ostby}}, \bibinfo {author} {\bibfnamefont {L.}~\bibnamefont {Yang}},\ and\
  \bibinfo {author} {\bibfnamefont {K.~J.}\ \bibnamefont {Vahala}},\ }\bibfield
   {title} {\bibinfo {title} {Ultralow-threshold {Yb$^{3+}$:SiO$_2$} glass
  laser fabricated by the solgel process},\ }\href
  {https://doi.org/10.1364/OL.32.002650} {\bibfield  {journal} {\bibinfo
  {journal} {Opt. Lett.}\ }\textbf {\bibinfo {volume} {32}},\ \bibinfo {pages}
  {2650} (\bibinfo {year} {2007})}\BibitemShut {NoStop}%
\bibitem [{\citenamefont {Herr}\ \emph {et~al.}(2018)\citenamefont {Herr},
  \citenamefont {Werner}, \citenamefont {Buse},\ and\ \citenamefont
  {Breunig}}]{Herr18}%
  \BibitemOpen
  \bibfield  {author} {\bibinfo {author} {\bibfnamefont {S.~J.}\ \bibnamefont
  {Herr}}, \bibinfo {author} {\bibfnamefont {C.~S.}\ \bibnamefont {Werner}},
  \bibinfo {author} {\bibfnamefont {K.}~\bibnamefont {Buse}},\ and\ \bibinfo
  {author} {\bibfnamefont {I.}~\bibnamefont {Breunig}},\ }\bibfield  {title}
  {\bibinfo {title} {Quasi-phase-matched self-pumped optical parametric
  oscillation in a micro-resonator},\ }\href
  {https://doi.org/10.1364/OE.26.010813} {\bibfield  {journal} {\bibinfo
  {journal} {Opt. Express}\ }\textbf {\bibinfo {volume} {26}},\ \bibinfo
  {pages} {10813} (\bibinfo {year} {2018})}\BibitemShut {NoStop}%
\bibitem [{\citenamefont {Lissillour}\ \emph {et~al.}(2001)\citenamefont
  {Lissillour}, \citenamefont {Messager}, \citenamefont {St\'{e}phan},\ and\
  \citenamefont {F\'{e}ron}}]{Lissillour01a}%
  \BibitemOpen
  \bibfield  {author} {\bibinfo {author} {\bibfnamefont {F.}~\bibnamefont
  {Lissillour}}, \bibinfo {author} {\bibfnamefont {D.}~\bibnamefont
  {Messager}}, \bibinfo {author} {\bibfnamefont {G.}~\bibnamefont
  {St\'{e}phan}},\ and\ \bibinfo {author} {\bibfnamefont {P.}~\bibnamefont
  {F\'{e}ron}},\ }\bibfield  {title} {\bibinfo {title} {Whispering-gallery-mode
  laser at 1.56$\mu$m excited by a fiber taper},\ }\href
  {https://doi.org/10.1364/OL.26.001051} {\bibfield  {journal} {\bibinfo
  {journal} {Opt. Lett.}\ }\textbf {\bibinfo {volume} {26}},\ \bibinfo {pages}
  {1051} (\bibinfo {year} {2001})}\BibitemShut {NoStop}%
\bibitem [{\citenamefont {Tabataba-Vakili}\ \emph {et~al.}(2018)\citenamefont
  {Tabataba-Vakili}, \citenamefont {Doyennette}, \citenamefont {Brimont},
  \citenamefont {Guillet}, \citenamefont {Rennesson}, \citenamefont
  {Frayssinet}, \citenamefont {Damilano}, \citenamefont {Duboz}, \citenamefont
  {Semond}, \citenamefont {Roland}, \citenamefont {El~Kurdi}, \citenamefont
  {Checoury}, \citenamefont {Sauvage}, \citenamefont {Gayral},\ and\
  \citenamefont {Boucaud}}]{Tabataba18}%
  \BibitemOpen
  \bibfield  {author} {\bibinfo {author} {\bibfnamefont {F.}~\bibnamefont
  {Tabataba-Vakili}}, \bibinfo {author} {\bibfnamefont {L.}~\bibnamefont
  {Doyennette}}, \bibinfo {author} {\bibfnamefont {C.}~\bibnamefont {Brimont}},
  \bibinfo {author} {\bibfnamefont {T.}~\bibnamefont {Guillet}}, \bibinfo
  {author} {\bibfnamefont {S.}~\bibnamefont {Rennesson}}, \bibinfo {author}
  {\bibfnamefont {E.}~\bibnamefont {Frayssinet}}, \bibinfo {author}
  {\bibfnamefont {B.}~\bibnamefont {Damilano}}, \bibinfo {author}
  {\bibfnamefont {J.-Y.}\ \bibnamefont {Duboz}}, \bibinfo {author}
  {\bibfnamefont {F.}~\bibnamefont {Semond}}, \bibinfo {author} {\bibfnamefont
  {I.}~\bibnamefont {Roland}}, \bibinfo {author} {\bibfnamefont
  {M.}~\bibnamefont {El~Kurdi}}, \bibinfo {author} {\bibfnamefont
  {X.}~\bibnamefont {Checoury}}, \bibinfo {author} {\bibfnamefont
  {S.}~\bibnamefont {Sauvage}}, \bibinfo {author} {\bibfnamefont
  {B.}~\bibnamefont {Gayral}},\ and\ \bibinfo {author} {\bibfnamefont
  {P.}~\bibnamefont {Boucaud}},\ }\bibfield  {title} {\bibinfo {title} {Blue
  microlasers integrated on a photonic platform on silicon},\ }\href
  {https://doi.org/10.1021/acsphotonics.8b00542} {\bibfield  {journal}
  {\bibinfo  {journal} {ACS Photonics}\ }\textbf {\bibinfo {volume} {5}},\
  \bibinfo {pages} {3643} (\bibinfo {year} {2018})}\BibitemShut {NoStop}%
\bibitem [{\citenamefont {Palma}\ \emph {et~al.}(2016)\citenamefont {Palma},
  \citenamefont {Falconi}, \citenamefont {Starecki}, \citenamefont {Nazabal},
  \citenamefont {Yano}, \citenamefont {Kishi}, \citenamefont {Kumagai},\ and\
  \citenamefont {Prudenzano}}]{Palma16}%
  \BibitemOpen
  \bibfield  {author} {\bibinfo {author} {\bibfnamefont {G.}~\bibnamefont
  {Palma}}, \bibinfo {author} {\bibfnamefont {M.~C.}\ \bibnamefont {Falconi}},
  \bibinfo {author} {\bibfnamefont {F.}~\bibnamefont {Starecki}}, \bibinfo
  {author} {\bibfnamefont {V.}~\bibnamefont {Nazabal}}, \bibinfo {author}
  {\bibfnamefont {T.}~\bibnamefont {Yano}}, \bibinfo {author} {\bibfnamefont
  {T.}~\bibnamefont {Kishi}}, \bibinfo {author} {\bibfnamefont
  {T.}~\bibnamefont {Kumagai}},\ and\ \bibinfo {author} {\bibfnamefont
  {F.}~\bibnamefont {Prudenzano}},\ }\bibfield  {title} {\bibinfo {title}
  {Novel double step approach for optical sensing via microsphere {WGM}
  resonance},\ }\href {https://doi.org/10.1364/OE.24.026956} {\bibfield
  {journal} {\bibinfo  {journal} {Opt. Express}\ }\textbf {\bibinfo {volume}
  {24}},\ \bibinfo {pages} {26956} (\bibinfo {year} {2016})}\BibitemShut
  {NoStop}%
\bibitem [{\citenamefont {Palma}\ \emph {et~al.}(2017)\citenamefont {Palma},
  \citenamefont {Falconi}, \citenamefont {Starecki}, \citenamefont {Nazabal},
  \citenamefont {Ari}, \citenamefont {Bodiou}, \citenamefont {Charrier},
  \citenamefont {Dumeige}, \citenamefont {Baudet},\ and\ \citenamefont
  {Prudenzano}}]{Palma17}%
  \BibitemOpen
  \bibfield  {author} {\bibinfo {author} {\bibfnamefont {G.}~\bibnamefont
  {Palma}}, \bibinfo {author} {\bibfnamefont {M.~C.}\ \bibnamefont {Falconi}},
  \bibinfo {author} {\bibfnamefont {F.}~\bibnamefont {Starecki}}, \bibinfo
  {author} {\bibfnamefont {V.}~\bibnamefont {Nazabal}}, \bibinfo {author}
  {\bibfnamefont {J.}~\bibnamefont {Ari}}, \bibinfo {author} {\bibfnamefont
  {L.}~\bibnamefont {Bodiou}}, \bibinfo {author} {\bibfnamefont
  {J.}~\bibnamefont {Charrier}}, \bibinfo {author} {\bibfnamefont
  {Y.}~\bibnamefont {Dumeige}}, \bibinfo {author} {\bibfnamefont
  {E.}~\bibnamefont {Baudet}},\ and\ \bibinfo {author} {\bibfnamefont
  {F.}~\bibnamefont {Prudenzano}},\ }\bibfield  {title} {\bibinfo {title}
  {Design of praseodymium-doped chalcogenide micro-disk emitting at 4.7
  $\mu$m},\ }\href {https://doi.org/10.1364/OE.25.007014} {\bibfield  {journal}
  {\bibinfo  {journal} {Opt. Express}\ }\textbf {\bibinfo {volume} {25}},\
  \bibinfo {pages} {7014} (\bibinfo {year} {2017})}\BibitemShut {NoStop}%
\bibitem [{\citenamefont {{Behzadi}}\ \emph {et~al.}(2017)\citenamefont
  {{Behzadi}}, \citenamefont {{Jain}},\ and\ \citenamefont
  {{Hossein-Zadeh}}}]{Behzadi17}%
  \BibitemOpen
  \bibfield  {author} {\bibinfo {author} {\bibfnamefont {B.}~\bibnamefont
  {{Behzadi}}}, \bibinfo {author} {\bibfnamefont {R.~K.}\ \bibnamefont
  {{Jain}}},\ and\ \bibinfo {author} {\bibfnamefont {M.}~\bibnamefont
  {{Hossein-Zadeh}}},\ }\bibfield  {title} {\bibinfo {title} {Spectral and
  modal properties of a mid-{IR} spherical microlaser},\ }\href
  {https://doi.org/10.1109/JQE.2017.2771423} {\bibfield  {journal} {\bibinfo
  {journal} {IEEE J. Quantum Electron.}\ }\textbf {\bibinfo {volume} {53}},\
  \bibinfo {pages} {1} (\bibinfo {year} {2017})}\BibitemShut {NoStop}%
\bibitem [{\citenamefont {Gerard}\ and\ \citenamefont
  {Gayral}(1999)}]{Gerard99}%
  \BibitemOpen
  \bibfield  {author} {\bibinfo {author} {\bibfnamefont {J.~M.}\ \bibnamefont
  {Gerard}}\ and\ \bibinfo {author} {\bibfnamefont {B.}~\bibnamefont
  {Gayral}},\ }\bibfield  {title} {\bibinfo {title} {Strong {Purcell effect for
  InAs} quantum boxes in three-dimensional solid-state microcavities},\ }\href
  {https://doi.org/10.1109/50.802999} {\bibfield  {journal} {\bibinfo
  {journal} {J. Lightwave Technol.}\ }\textbf {\bibinfo {volume} {17}},\
  \bibinfo {pages} {2089} (\bibinfo {year} {1999})}\BibitemShut {NoStop}%
\bibitem [{\citenamefont {Huet}\ \emph {et~al.}(2016)\citenamefont {Huet},
  \citenamefont {Rasoloniaina}, \citenamefont {Guillem\'e}, \citenamefont
  {Rochard}, \citenamefont {F\'eron}, \citenamefont {Mortier}, \citenamefont
  {Levenson}, \citenamefont {Bencheikh}, \citenamefont {Yacomotti},\ and\
  \citenamefont {Dumeige}}]{Huet2016}%
  \BibitemOpen
  \bibfield  {author} {\bibinfo {author} {\bibfnamefont {V.}~\bibnamefont
  {Huet}}, \bibinfo {author} {\bibfnamefont {A.}~\bibnamefont {Rasoloniaina}},
  \bibinfo {author} {\bibfnamefont {P.}~\bibnamefont {Guillem\'e}}, \bibinfo
  {author} {\bibfnamefont {P.}~\bibnamefont {Rochard}}, \bibinfo {author}
  {\bibfnamefont {P.}~\bibnamefont {F\'eron}}, \bibinfo {author} {\bibfnamefont
  {M.}~\bibnamefont {Mortier}}, \bibinfo {author} {\bibfnamefont
  {A.}~\bibnamefont {Levenson}}, \bibinfo {author} {\bibfnamefont
  {K.}~\bibnamefont {Bencheikh}}, \bibinfo {author} {\bibfnamefont
  {A.}~\bibnamefont {Yacomotti}},\ and\ \bibinfo {author} {\bibfnamefont
  {Y.}~\bibnamefont {Dumeige}},\ }\bibfield  {title} {\bibinfo {title}
  {Millisecond photon lifetime in a slow-light microcavity},\ }\href
  {https://doi.org/10.1103/PhysRevLett.116.133902} {\bibfield  {journal}
  {\bibinfo  {journal} {Phys. Rev. Lett.}\ }\textbf {\bibinfo {volume} {116}},\
  \bibinfo {pages} {133902} (\bibinfo {year} {2016})}\BibitemShut {NoStop}%
\bibitem [{\citenamefont {Ciminelli}\ \emph {et~al.}(2010)\citenamefont
  {Ciminelli}, \citenamefont {Dell'Olio}, \citenamefont {Campanella},\ and\
  \citenamefont {Armenise}}]{Ciminelli10}%
  \BibitemOpen
  \bibfield  {author} {\bibinfo {author} {\bibfnamefont {C.}~\bibnamefont
  {Ciminelli}}, \bibinfo {author} {\bibfnamefont {F.}~\bibnamefont
  {Dell'Olio}}, \bibinfo {author} {\bibfnamefont {C.~E.}\ \bibnamefont
  {Campanella}},\ and\ \bibinfo {author} {\bibfnamefont {M.~N.}\ \bibnamefont
  {Armenise}},\ }\bibfield  {title} {\bibinfo {title} {Photonic technologies
  for angular velocity sensing},\ }\href {https://doi.org/10.1364/AOP.2.000370}
  {\bibfield  {journal} {\bibinfo  {journal} {Adv. Opt. Photonics}\ }\textbf
  {\bibinfo {volume} {2}},\ \bibinfo {pages} {370} (\bibinfo {year}
  {2010})}\BibitemShut {NoStop}%
\bibitem [{\citenamefont {Ward}\ and\ \citenamefont {Benson}(2011)}]{Ward11}%
  \BibitemOpen
  \bibfield  {author} {\bibinfo {author} {\bibfnamefont {J.}~\bibnamefont
  {Ward}}\ and\ \bibinfo {author} {\bibfnamefont {O.}~\bibnamefont {Benson}},\
  }\bibfield  {title} {\bibinfo {title} {{WGM} microresonators: sensing, lasing
  and fundamental optics with microspheres},\ }\href
  {https://doi.org/10.1002/lpor.201000025} {\bibfield  {journal} {\bibinfo
  {journal} {Laser Photonics Rev.}\ }\textbf {\bibinfo {volume} {5}},\ \bibinfo
  {pages} {553} (\bibinfo {year} {2011})}\BibitemShut {NoStop}%
\bibitem [{\citenamefont {Ren}\ \emph {et~al.}(2017)\citenamefont {Ren},
  \citenamefont {Hodaei}, \citenamefont {Harari}, \citenamefont {Hassan},
  \citenamefont {Chow}, \citenamefont {Soltani}, \citenamefont
  {Christodoulides},\ and\ \citenamefont {Khajavikhan}}]{Ren17}%
  \BibitemOpen
  \bibfield  {author} {\bibinfo {author} {\bibfnamefont {J.}~\bibnamefont
  {Ren}}, \bibinfo {author} {\bibfnamefont {H.}~\bibnamefont {Hodaei}},
  \bibinfo {author} {\bibfnamefont {G.}~\bibnamefont {Harari}}, \bibinfo
  {author} {\bibfnamefont {A.~U.}\ \bibnamefont {Hassan}}, \bibinfo {author}
  {\bibfnamefont {W.}~\bibnamefont {Chow}}, \bibinfo {author} {\bibfnamefont
  {M.}~\bibnamefont {Soltani}}, \bibinfo {author} {\bibfnamefont
  {D.}~\bibnamefont {Christodoulides}},\ and\ \bibinfo {author} {\bibfnamefont
  {M.}~\bibnamefont {Khajavikhan}},\ }\bibfield  {title} {\bibinfo {title}
  {Ultrasensitive micro-scale parity-time-symmetric ring laser gyroscope},\
  }\href {https://doi.org/10.1364/OL.42.001556} {\bibfield  {journal} {\bibinfo
   {journal} {Opt. Lett.}\ }\textbf {\bibinfo {volume} {42}},\ \bibinfo {pages}
  {1556} (\bibinfo {year} {2017})}\BibitemShut {NoStop}%
\bibitem [{\citenamefont {Liu}\ \emph {et~al.}(2010)\citenamefont {Liu},
  \citenamefont {Kumar}, \citenamefont {Huybrechts}, \citenamefont {Spuesens},
  \citenamefont {Roelkens}, \citenamefont {Geluk}, \citenamefont {de~Vries},
  \citenamefont {Regreny}, \citenamefont {van Thourhout}, \citenamefont
  {Baets},\ and\ \citenamefont {Morthier}}]{Liu10}%
  \BibitemOpen
  \bibfield  {author} {\bibinfo {author} {\bibfnamefont {L.}~\bibnamefont
  {Liu}}, \bibinfo {author} {\bibfnamefont {R.}~\bibnamefont {Kumar}}, \bibinfo
  {author} {\bibfnamefont {K.}~\bibnamefont {Huybrechts}}, \bibinfo {author}
  {\bibfnamefont {T.}~\bibnamefont {Spuesens}}, \bibinfo {author}
  {\bibfnamefont {G.}~\bibnamefont {Roelkens}}, \bibinfo {author}
  {\bibfnamefont {E.-J.}\ \bibnamefont {Geluk}}, \bibinfo {author}
  {\bibfnamefont {T.}~\bibnamefont {de~Vries}}, \bibinfo {author}
  {\bibfnamefont {P.}~\bibnamefont {Regreny}}, \bibinfo {author} {\bibfnamefont
  {D.}~\bibnamefont {van Thourhout}}, \bibinfo {author} {\bibfnamefont
  {R.}~\bibnamefont {Baets}},\ and\ \bibinfo {author} {\bibfnamefont
  {G.}~\bibnamefont {Morthier}},\ }\href@noop {} {\bibfield  {journal}
  {\bibinfo  {journal} {Nature Photonics}\ }\textbf {\bibinfo {volume} {4}},\
  \bibinfo {pages} {182} (\bibinfo {year} {2010})}\BibitemShut {NoStop}%
\bibitem [{\citenamefont {Xiao}\ \emph {et~al.}(2010)\citenamefont {Xiao},
  \citenamefont {Trebaol}, \citenamefont {Dumeige}, \citenamefont {Cai},
  \citenamefont {Mortier},\ and\ \citenamefont {Feron}}]{Xiao10}%
  \BibitemOpen
  \bibfield  {author} {\bibinfo {author} {\bibfnamefont {L.}~\bibnamefont
  {Xiao}}, \bibinfo {author} {\bibfnamefont {S.}~\bibnamefont {Trebaol}},
  \bibinfo {author} {\bibfnamefont {Y.}~\bibnamefont {Dumeige}}, \bibinfo
  {author} {\bibfnamefont {Z.}~\bibnamefont {Cai}}, \bibinfo {author}
  {\bibfnamefont {M.}~\bibnamefont {Mortier}},\ and\ \bibinfo {author}
  {\bibfnamefont {P.}~\bibnamefont {Feron}},\ }\bibfield  {title} {\bibinfo
  {title} {Miniaturized optical microwave source using a dual-wavelength
  whispering gallery mode laser},\ }\href
  {https://doi.org/10.1109/LPT.2010.2040269} {\bibfield  {journal} {\bibinfo
  {journal} {IEEE Photonics Technol. Lett.}\ }\textbf {\bibinfo {volume}
  {22}},\ \bibinfo {pages} {559} (\bibinfo {year} {2010})}\BibitemShut
  {NoStop}%
\bibitem [{\citenamefont {{F. Lissillour}}\ \emph {et~al.}(2001)\citenamefont
  {{F. Lissillour}}, \citenamefont {{R. Gabet}}, \citenamefont {{P.
  F\'{e}ron}}, \citenamefont {{P. Besnard}},\ and\ \citenamefont {{G.
  St\'{e}phan}}}]{Lissillour01}%
  \BibitemOpen
  \bibfield  {author} {\bibinfo {author} {\bibnamefont {{F. Lissillour}}},
  \bibinfo {author} {\bibnamefont {{R. Gabet}}}, \bibinfo {author}
  {\bibnamefont {{P. F\'{e}ron}}}, \bibinfo {author} {\bibnamefont {{P.
  Besnard}}},\ and\ \bibinfo {author} {\bibnamefont {{G. St\'{e}phan}}},\
  }\bibfield  {title} {\bibinfo {title} {Linewidth narrowing of a {DFB}
  semiconductor laser at $1.55\mu{\rm m}$ by optical injection of an {Er:ZBLAN}
  microspherical laser},\ }\href {https://doi.org/10.1209/epl/i2001-00443-7}
  {\bibfield  {journal} {\bibinfo  {journal} {EPL (Europhysics Letters)}\
  }\textbf {\bibinfo {volume} {55}},\ \bibinfo {pages} {499} (\bibinfo {year}
  {2001})}\BibitemShut {NoStop}%
\bibitem [{\citenamefont {He}\ \emph {et~al.}(2010{\natexlab{a}})\citenamefont
  {He}, \citenamefont {Ozdemir}, \citenamefont {Xiao},\ and\ \citenamefont
  {Yang}}]{He10}%
  \BibitemOpen
  \bibfield  {author} {\bibinfo {author} {\bibfnamefont {L.}~\bibnamefont
  {He}}, \bibinfo {author} {\bibfnamefont {S.~K.}\ \bibnamefont {Ozdemir}},
  \bibinfo {author} {\bibfnamefont {Y.~F.}\ \bibnamefont {Xiao}},\ and\
  \bibinfo {author} {\bibfnamefont {L.}~\bibnamefont {Yang}},\ }\bibfield
  {title} {\bibinfo {title} {Gain-induced evolution of mode splitting spectra
  in a high-{Q} active microresonator},\ }\href
  {https://doi.org/10.1109/JQE.2010.2055549} {\bibfield  {journal} {\bibinfo
  {journal} {IEEE J. Quantum Electron.}\ }\textbf {\bibinfo {volume} {46}},\
  \bibinfo {pages} {1626} (\bibinfo {year} {2010}{\natexlab{a}})}\BibitemShut
  {NoStop}%
\bibitem [{\citenamefont {Schwartz}\ \emph {et~al.}(2006)\citenamefont
  {Schwartz}, \citenamefont {Feugnet}, \citenamefont {Bouyer}, \citenamefont
  {Lariontsev}, \citenamefont {Aspect},\ and\ \citenamefont
  {Pocholle}}]{Schwartz06}%
  \BibitemOpen
  \bibfield  {author} {\bibinfo {author} {\bibfnamefont {S.}~\bibnamefont
  {Schwartz}}, \bibinfo {author} {\bibfnamefont {G.}~\bibnamefont {Feugnet}},
  \bibinfo {author} {\bibfnamefont {P.}~\bibnamefont {Bouyer}}, \bibinfo
  {author} {\bibfnamefont {E.}~\bibnamefont {Lariontsev}}, \bibinfo {author}
  {\bibfnamefont {A.}~\bibnamefont {Aspect}},\ and\ \bibinfo {author}
  {\bibfnamefont {J.-P.}\ \bibnamefont {Pocholle}},\ }\bibfield  {title}
  {\bibinfo {title} {Mode-coupling control in resonant devices: Application to
  solid-state ring lasers},\ }\href
  {http://journals.aps.org/prl/pdf/10.1103/PhysRevLett.97.093902} {\bibfield
  {journal} {\bibinfo  {journal} {Phys. Rev. Lett.}\ }\textbf {\bibinfo
  {volume} {97}},\ \bibinfo {pages} {093902} (\bibinfo {year}
  {2006})}\BibitemShut {NoStop}%
\bibitem [{\citenamefont {Schwartz}\ \emph {et~al.}(2007)\citenamefont
  {Schwartz}, \citenamefont {Feugnet}, \citenamefont {Lariontsev},\ and\
  \citenamefont {Pocholle}}]{Schwartz07}%
  \BibitemOpen
  \bibfield  {author} {\bibinfo {author} {\bibfnamefont {S.}~\bibnamefont
  {Schwartz}}, \bibinfo {author} {\bibfnamefont {G.}~\bibnamefont {Feugnet}},
  \bibinfo {author} {\bibfnamefont {E.}~\bibnamefont {Lariontsev}},\ and\
  \bibinfo {author} {\bibfnamefont {J.-P.}\ \bibnamefont {Pocholle}},\
  }\bibfield  {title} {\bibinfo {title} {Oscillation regimes of a solid-state
  ring laser with active beat-note stabilization: From a chaotic device to a
  ring-laser gyroscope},\ }\href
  {http://journals.aps.org/pra/pdf/10.1103/PhysRevA.76.023807} {\bibfield
  {journal} {\bibinfo  {journal} {Phys. Rev. A}\ }\textbf {\bibinfo {volume}
  {76}},\ \bibinfo {pages} {023807} (\bibinfo {year} {2007})}\BibitemShut
  {NoStop}%
\bibitem [{\citenamefont {Schwartz}\ \emph {et~al.}(2008)\citenamefont
  {Schwartz}, \citenamefont {Gutty}, \citenamefont {Feugnet}, \citenamefont
  {Bouyer},\ and\ \citenamefont {Pocholle}}]{Schwartz08}%
  \BibitemOpen
  \bibfield  {author} {\bibinfo {author} {\bibfnamefont {S.}~\bibnamefont
  {Schwartz}}, \bibinfo {author} {\bibfnamefont {F.}~\bibnamefont {Gutty}},
  \bibinfo {author} {\bibfnamefont {G.}~\bibnamefont {Feugnet}}, \bibinfo
  {author} {\bibfnamefont {P.}~\bibnamefont {Bouyer}},\ and\ \bibinfo {author}
  {\bibfnamefont {J.-P.}\ \bibnamefont {Pocholle}},\ }\bibfield  {title}
  {\bibinfo {title} {Suppression of nonlinear interactions in resonant
  macroscopic quantum devices: The example of the solid-state ring laser
  gyroscope},\ }\href
  {http://journals.aps.org/prl/pdf/10.1103/PhysRevLett.100.183901} {\bibfield
  {journal} {\bibinfo  {journal} {Phys. Rev Lett}\ }\textbf {\bibinfo {volume}
  {100}},\ \bibinfo {pages} {183901} (\bibinfo {year} {2008})}\BibitemShut
  {NoStop}%
\bibitem [{\citenamefont {Efanova}\ and\ \citenamefont
  {Lariontsev}(1968)}]{efanova1968interaction}%
  \BibitemOpen
  \bibfield  {author} {\bibinfo {author} {\bibfnamefont {I.}~\bibnamefont
  {Efanova}}\ and\ \bibinfo {author} {\bibfnamefont {E.}~\bibnamefont
  {Lariontsev}},\ }\bibfield  {title} {\bibinfo {title} {Interaction of
  oppositely directed waves in a solid state ring laser},\ }\href
  {http://jetp.ac.ru/cgi-bin/dn/e_028_04_0802.pdf} {\bibfield  {journal}
  {\bibinfo  {journal} {Soviet Phys. JETP}\ }\textbf {\bibinfo {volume} {28}},\
  \bibinfo {pages} {802} (\bibinfo {year} {1968})}\BibitemShut {NoStop}%
\bibitem [{\citenamefont {Khandokhin}\ and\ \citenamefont
  {Khanin}(1985)}]{Khandokhin85}%
  \BibitemOpen
  \bibfield  {author} {\bibinfo {author} {\bibfnamefont {P.~A.}\ \bibnamefont
  {Khandokhin}}\ and\ \bibinfo {author} {\bibfnamefont {Y.~I.}\ \bibnamefont
  {Khanin}},\ }\bibfield  {title} {\bibinfo {title} {Instabilities in a
  solid-state ring laser},\ }\href {https://doi.org/10.1364/JOSAB.2.000226}
  {\bibfield  {journal} {\bibinfo  {journal} {J. Opt. Soc. Am. B}\ }\textbf
  {\bibinfo {volume} {2}},\ \bibinfo {pages} {226} (\bibinfo {year}
  {1985})}\BibitemShut {NoStop}%
\bibitem [{\citenamefont {Weiss}\ \emph {et~al.}(1995)\citenamefont {Weiss},
  \citenamefont {Sandoghdar}, \citenamefont {Hare}, \citenamefont
  {Lefevre-Seguin}, \citenamefont {Raimond},\ and\ \citenamefont
  {Haroche}}]{Weiss95}%
  \BibitemOpen
  \bibfield  {author} {\bibinfo {author} {\bibfnamefont {D.}~\bibnamefont
  {Weiss}}, \bibinfo {author} {\bibfnamefont {V.}~\bibnamefont {Sandoghdar}},
  \bibinfo {author} {\bibfnamefont {J.}~\bibnamefont {Hare}}, \bibinfo {author}
  {\bibfnamefont {V.}~\bibnamefont {Lefevre-Seguin}}, \bibinfo {author}
  {\bibfnamefont {J.-M.}\ \bibnamefont {Raimond}},\ and\ \bibinfo {author}
  {\bibfnamefont {S.}~\bibnamefont {Haroche}},\ }\bibfield  {title} {\bibinfo
  {title} {Splitting of high-${Q}$ {M}ie modes induced by light backscattering
  in silica microspheres},\ }\href@noop {} {\bibfield  {journal} {\bibinfo
  {journal} {Opt. Lett.}\ }\textbf {\bibinfo {volume} {20}},\ \bibinfo {pages}
  {1835} (\bibinfo {year} {1995})}\BibitemShut {NoStop}%
\bibitem [{\citenamefont {Kippenberg}\ \emph {et~al.}(2009)\citenamefont
  {Kippenberg}, \citenamefont {Tchebotareva}, \citenamefont {Kalkman},
  \citenamefont {Polman},\ and\ \citenamefont {Vahala}}]{Kippenberg2009}%
  \BibitemOpen
  \bibfield  {author} {\bibinfo {author} {\bibfnamefont {T.~J.}\ \bibnamefont
  {Kippenberg}}, \bibinfo {author} {\bibfnamefont {A.~L.}\ \bibnamefont
  {Tchebotareva}}, \bibinfo {author} {\bibfnamefont {J.}~\bibnamefont
  {Kalkman}}, \bibinfo {author} {\bibfnamefont {A.}~\bibnamefont {Polman}},\
  and\ \bibinfo {author} {\bibfnamefont {K.~J.}\ \bibnamefont {Vahala}},\
  }\bibfield  {title} {\bibinfo {title} {Purcell-factor-enhanced scattering
  from si nanocrystals in an optical microcavity},\ }\href
  {https://doi.org/10.1103/PhysRevLett.103.027406} {\bibfield  {journal}
  {\bibinfo  {journal} {Phys. Rev. Lett.}\ }\textbf {\bibinfo {volume} {103}},\
  \bibinfo {pages} {027406} (\bibinfo {year} {2009})}\BibitemShut {NoStop}%
\bibitem [{\citenamefont {Trebaol}\ \emph {et~al.}(2010)\citenamefont
  {Trebaol}, \citenamefont {Dumeige},\ and\ \citenamefont
  {F\'eron}}]{Trebaol2010}%
  \BibitemOpen
  \bibfield  {author} {\bibinfo {author} {\bibfnamefont {S.}~\bibnamefont
  {Trebaol}}, \bibinfo {author} {\bibfnamefont {Y.}~\bibnamefont {Dumeige}},\
  and\ \bibinfo {author} {\bibfnamefont {P.}~\bibnamefont {F\'eron}},\
  }\bibfield  {title} {\bibinfo {title} {Ringing phenomenon in coupled
  cavities: Application to modal coupling in {W}hispering-{G}allery-{M}ode
  resonators},\ }\href {https://doi.org/10.1103/PhysRevA.81.043828} {\bibfield
  {journal} {\bibinfo  {journal} {Phys. Rev. A}\ }\textbf {\bibinfo {volume}
  {81}},\ \bibinfo {pages} {043828} (\bibinfo {year} {2010})}\BibitemShut
  {NoStop}%
\bibitem [{\citenamefont {{Sorel}}\ \emph {et~al.}(2003)\citenamefont
  {{Sorel}}, \citenamefont {{Giuliani}}, \citenamefont {{Scire}}, \citenamefont
  {{Miglierina}}, \citenamefont {{Donati}},\ and\ \citenamefont
  {{Laybourn}}}]{Sorel03}%
  \BibitemOpen
  \bibfield  {author} {\bibinfo {author} {\bibfnamefont {M.}~\bibnamefont
  {{Sorel}}}, \bibinfo {author} {\bibfnamefont {G.}~\bibnamefont {{Giuliani}}},
  \bibinfo {author} {\bibfnamefont {A.}~\bibnamefont {{Scire}}}, \bibinfo
  {author} {\bibfnamefont {R.}~\bibnamefont {{Miglierina}}}, \bibinfo {author}
  {\bibfnamefont {S.}~\bibnamefont {{Donati}}},\ and\ \bibinfo {author}
  {\bibfnamefont {P.~J.~R.}\ \bibnamefont {{Laybourn}}},\ }\bibfield  {title}
  {\bibinfo {title} {Operating regimes of {GaAs-AlGaAs} semiconductor ring
  lasers: experiment and model},\ }\href
  {https://doi.org/10.1109/JQE.2003.817585} {\bibfield  {journal} {\bibinfo
  {journal} {IEEE J. Quantum Electron.}\ }\textbf {\bibinfo {volume} {39}},\
  \bibinfo {pages} {1187} (\bibinfo {year} {2003})}\BibitemShut {NoStop}%
\bibitem [{\citenamefont {{Giuliani}}\ \emph {et~al.}(2005)\citenamefont
  {{Giuliani}}, \citenamefont {{Miglierina}}, \citenamefont {{Sorel}},\ and\
  \citenamefont {{Scire}}}]{Giuliani05}%
  \BibitemOpen
  \bibfield  {author} {\bibinfo {author} {\bibfnamefont {G.}~\bibnamefont
  {{Giuliani}}}, \bibinfo {author} {\bibfnamefont {R.}~\bibnamefont
  {{Miglierina}}}, \bibinfo {author} {\bibfnamefont {M.}~\bibnamefont
  {{Sorel}}},\ and\ \bibinfo {author} {\bibfnamefont {A.}~\bibnamefont
  {{Scire}}},\ }\bibfield  {title} {\bibinfo {title} {Linewidth,
  autocorrelation, and cross-correlation measurements of counterpropagating
  modes in {GaAs-AlGaAs} semiconductor ring lasers},\ }\href
  {https://doi.org/10.1109/JSTQE.2005.854144} {\bibfield  {journal} {\bibinfo
  {journal} {IEEE J. Sel. Topics Quantum Electron.}\ }\textbf {\bibinfo
  {volume} {11}},\ \bibinfo {pages} {1187} (\bibinfo {year}
  {2005})}\BibitemShut {NoStop}%
\bibitem [{\citenamefont {{Javaloyes}}\ and\ \citenamefont
  {{Balle}}(2009)}]{Javaloyes09}%
  \BibitemOpen
  \bibfield  {author} {\bibinfo {author} {\bibfnamefont {J.}~\bibnamefont
  {{Javaloyes}}}\ and\ \bibinfo {author} {\bibfnamefont {S.}~\bibnamefont
  {{Balle}}},\ }\bibfield  {title} {\bibinfo {title} {Emission directionality
  of semiconductor ring lasers: A traveling-wave description},\ }\href
  {https://doi.org/10.1109/JQE.2009.2014079} {\bibfield  {journal} {\bibinfo
  {journal} {IEEE J. Quantum Electron.}\ }\textbf {\bibinfo {volume} {45}},\
  \bibinfo {pages} {431} (\bibinfo {year} {2009})}\BibitemShut {NoStop}%
\bibitem [{\citenamefont {He}\ \emph {et~al.}(2010{\natexlab{b}})\citenamefont
  {He}, \citenamefont {\"Ozdemir}, \citenamefont {Zhu},\ and\ \citenamefont
  {Yang}}]{He10_pra}%
  \BibitemOpen
  \bibfield  {author} {\bibinfo {author} {\bibfnamefont {L.}~\bibnamefont
  {He}}, \bibinfo {author} {\bibfnamefont {S.~K.}\ \bibnamefont {\"Ozdemir}},
  \bibinfo {author} {\bibfnamefont {J.}~\bibnamefont {Zhu}},\ and\ \bibinfo
  {author} {\bibfnamefont {L.}~\bibnamefont {Yang}},\ }\bibfield  {title}
  {\bibinfo {title} {Ultrasensitive detection of mode splitting in active
  optical microcavities},\ }\href {https://doi.org/10.1103/PhysRevA.82.053810}
  {\bibfield  {journal} {\bibinfo  {journal} {Phys. Rev. A}\ }\textbf {\bibinfo
  {volume} {82}},\ \bibinfo {pages} {053810} (\bibinfo {year}
  {2010}{\natexlab{b}})}\BibitemShut {NoStop}%
\bibitem [{\citenamefont {Risti\'{c}}\ \emph {et~al.}(2016)\citenamefont
  {Risti\'{c}}, \citenamefont {Berneschi}, \citenamefont {Camerini},
  \citenamefont {Farnesi}, \citenamefont {Pelli}, \citenamefont {Trono},
  \citenamefont {Chiappini}, \citenamefont {Chiasera}, \citenamefont {Ferrari},
  \citenamefont {Lukowiak}, \citenamefont {Dumeige}, \citenamefont {F\'{e}ron},
  \citenamefont {Righini}, \citenamefont {Soria},\ and\ \citenamefont
  {Conti}}]{Ristic16}%
  \BibitemOpen
  \bibfield  {author} {\bibinfo {author} {\bibfnamefont {D.}~\bibnamefont
  {Risti\'{c}}}, \bibinfo {author} {\bibfnamefont {S.}~\bibnamefont
  {Berneschi}}, \bibinfo {author} {\bibfnamefont {M.}~\bibnamefont {Camerini}},
  \bibinfo {author} {\bibfnamefont {D.}~\bibnamefont {Farnesi}}, \bibinfo
  {author} {\bibfnamefont {S.}~\bibnamefont {Pelli}}, \bibinfo {author}
  {\bibfnamefont {C.}~\bibnamefont {Trono}}, \bibinfo {author} {\bibfnamefont
  {A.}~\bibnamefont {Chiappini}}, \bibinfo {author} {\bibfnamefont
  {A.}~\bibnamefont {Chiasera}}, \bibinfo {author} {\bibfnamefont
  {M.}~\bibnamefont {Ferrari}}, \bibinfo {author} {\bibfnamefont
  {A.}~\bibnamefont {Lukowiak}}, \bibinfo {author} {\bibfnamefont
  {Y.}~\bibnamefont {Dumeige}}, \bibinfo {author} {\bibfnamefont
  {P.}~\bibnamefont {F\'{e}ron}}, \bibinfo {author} {\bibfnamefont
  {G.}~\bibnamefont {Righini}}, \bibinfo {author} {\bibfnamefont
  {S.}~\bibnamefont {Soria}},\ and\ \bibinfo {author} {\bibfnamefont {G.~N.}\
  \bibnamefont {Conti}},\ }\bibfield  {title} {\bibinfo {title}
  {Photoluminescence and lasing in whispering gallery mode glass microspherical
  resonators},\ }\href
  {https://doi.org/https://doi.org/10.1016/j.jlumin.2015.10.050} {\bibfield
  {journal} {\bibinfo  {journal} {J. Lumin.}\ }\textbf {\bibinfo {volume}
  {170}},\ \bibinfo {pages} {755 } (\bibinfo {year} {2016})}\BibitemShut
  {NoStop}%
\bibitem [{\citenamefont {Rasoloniaina}\ \emph {et~al.}(2014)\citenamefont
  {Rasoloniaina}, \citenamefont {Huet}, \citenamefont {Nguy{\^{e}}n},
  \citenamefont {Cren}, \citenamefont {Mortier}, \citenamefont {Michely},
  \citenamefont {Dumeige},\ and\ \citenamefont
  {F{\'{e}}ron}}]{Rasoloniaina2014}%
  \BibitemOpen
  \bibfield  {author} {\bibinfo {author} {\bibfnamefont {A.}~\bibnamefont
  {Rasoloniaina}}, \bibinfo {author} {\bibfnamefont {V.}~\bibnamefont {Huet}},
  \bibinfo {author} {\bibfnamefont {T.~K.~N.}\ \bibnamefont {Nguy{\^{e}}n}},
  \bibinfo {author} {\bibfnamefont {E.~L.}\ \bibnamefont {Cren}}, \bibinfo
  {author} {\bibfnamefont {M.}~\bibnamefont {Mortier}}, \bibinfo {author}
  {\bibfnamefont {L.}~\bibnamefont {Michely}}, \bibinfo {author} {\bibfnamefont
  {Y.}~\bibnamefont {Dumeige}},\ and\ \bibinfo {author} {\bibfnamefont
  {P.}~\bibnamefont {F{\'{e}}ron}},\ }\bibfield  {title} {\bibinfo {title}
  {Controling the coupling properties of active ultrahigh-{Q} {WGM}
  microcavities from undercoupling to selective amplification},\ }\href@noop {}
  {\bibfield  {journal} {\bibinfo  {journal} {Sci. Rep.}\ }\textbf {\bibinfo
  {volume} {4}},\ \bibinfo {pages} {4023} (\bibinfo {year} {2014})}\BibitemShut
  {NoStop}%
\bibitem [{\citenamefont {Siegman}(1986)}]{siegman1986lasers}%
  \BibitemOpen
  \bibfield  {author} {\bibinfo {author} {\bibfnamefont {A.~E.}\ \bibnamefont
  {Siegman}},\ }\href@noop {} {\emph {\bibinfo {title} {Lasers}}}\ (\bibinfo
  {publisher} {University Science Books, Mill Valley, Calif},\ \bibinfo {year}
  {1986})\BibitemShut {NoStop}%
\bibitem [{\citenamefont {Schwartz}(2006)}]{schwartz2006gyrolaser}%
  \BibitemOpen
  \bibfield  {author} {\bibinfo {author} {\bibfnamefont {S.}~\bibnamefont
  {Schwartz}},\ }\emph {\bibinfo {title} {Gyrolaser {\`a} {\'e}tat solide.
  Application des lasers {\`a} atomes {\`a} la gyrom{\'e}trie.}},\ \href@noop
  {} {\bibinfo {type} {Th\`{e}se de doctorat}},\ \bibinfo  {school} {Ecole
  Polytechnique X} (\bibinfo {year} {2006})\BibitemShut {NoStop}%
\bibitem [{\citenamefont {Perevedentseva}\ \emph {et~al.}(1980)\citenamefont
  {Perevedentseva}, \citenamefont {Khandokhin},\ and\ \citenamefont
  {Khanin}}]{perevedentseva1980theory}%
  \BibitemOpen
  \bibfield  {author} {\bibinfo {author} {\bibfnamefont {G.}~\bibnamefont
  {Perevedentseva}}, \bibinfo {author} {\bibfnamefont {P.~A.}\ \bibnamefont
  {Khandokhin}},\ and\ \bibinfo {author} {\bibfnamefont {Y.~I.}\ \bibnamefont
  {Khanin}},\ }\bibfield  {title} {\bibinfo {title} {Theory of a
  single-frequency solid-state ring laser},\ }\href@noop {} {\bibfield
  {journal} {\bibinfo  {journal} {Soviet J. Quantum Electron.}\ }\textbf
  {\bibinfo {volume} {10}},\ \bibinfo {pages} {71} (\bibinfo {year}
  {1980})}\BibitemShut {NoStop}%
\bibitem [{\citenamefont {Ceppe}(2018)}]{Ceppe18}%
  \BibitemOpen
  \bibfield  {author} {\bibinfo {author} {\bibfnamefont {J.-B.}\ \bibnamefont
  {Ceppe}},\ }\emph {\bibinfo {title} {El\'{e}ments de dynamique du laser pour
  l'\'{e}laboration d'une source micro-onde miniaturis\'{e}e sur la base de la
  double \'{e}mission monomode d'un laser \`{a} modes de galerie}},\ \href@noop
  {} {\bibinfo {type} {Th\`{e}se de doctorat}},\ \bibinfo  {school}
  {Universit\'{e} de Rennes 1} (\bibinfo {year} {2018})\BibitemShut {NoStop}%
\bibitem [{\citenamefont {Ceppe}\ \emph {et~al.}(2017)\citenamefont {Ceppe},
  \citenamefont {Mortier}, \citenamefont {F\'{e}ron},\ and\ \citenamefont
  {Dumeige}}]{Ceppe17}%
  \BibitemOpen
  \bibfield  {author} {\bibinfo {author} {\bibfnamefont {J.-B.}\ \bibnamefont
  {Ceppe}}, \bibinfo {author} {\bibfnamefont {M.}~\bibnamefont {Mortier}},
  \bibinfo {author} {\bibfnamefont {P.}~\bibnamefont {F\'{e}ron}},\ and\
  \bibinfo {author} {\bibfnamefont {Y.}~\bibnamefont {Dumeige}},\ }\bibfield
  {title} {\bibinfo {title} {Theoretical and experimental analysis of rare
  earth whispering gallery mode laser relative intensity noise},\ }\href
  {https://doi.org/10.1364/OE.25.032732} {\bibfield  {journal} {\bibinfo
  {journal} {Opt. Express}\ }\textbf {\bibinfo {volume} {25}},\ \bibinfo
  {pages} {32732} (\bibinfo {year} {2017})}\BibitemShut {NoStop}%
\bibitem [{\citenamefont {Dumeige}\ \emph {et~al.}(2008)\citenamefont
  {Dumeige}, \citenamefont {Trebaol}, \citenamefont {Ghi\c{s}a}, \citenamefont
  {Nguy\^{e}n}, \citenamefont {Tavernier},\ and\ \citenamefont
  {F\'{e}ron}}]{Dumeige08}%
  \BibitemOpen
  \bibfield  {author} {\bibinfo {author} {\bibfnamefont {Y.}~\bibnamefont
  {Dumeige}}, \bibinfo {author} {\bibfnamefont {S.}~\bibnamefont {Trebaol}},
  \bibinfo {author} {\bibfnamefont {L.}~\bibnamefont {Ghi\c{s}a}}, \bibinfo
  {author} {\bibfnamefont {T.~K.~N.}\ \bibnamefont {Nguy\^{e}n}}, \bibinfo
  {author} {\bibfnamefont {H.}~\bibnamefont {Tavernier}},\ and\ \bibinfo
  {author} {\bibfnamefont {P.}~\bibnamefont {F\'{e}ron}},\ }\bibfield  {title}
  {\bibinfo {title} {Determination of coupling regime of high-Q resonators and
  optical gain of highly selective amplifiers},\ }\href
  {https://doi.org/10.1364/JOSAB.25.002073} {\bibfield  {journal} {\bibinfo
  {journal} {J. Opt. Soc. Am. B}\ }\textbf {\bibinfo {volume} {25}},\ \bibinfo
  {pages} {2073} (\bibinfo {year} {2008})}\BibitemShut {NoStop}%
\end{thebibliography}
\end{document}